\documentclass[conference]{IEEEtran}
\usepackage{amsfonts}
\usepackage{amsmath, amssymb, bm}
\usepackage{dsfont}
\usepackage{graphicx}
\usepackage{cite}
\usepackage{bigstrut}
\usepackage{float}
\usepackage{balance}
\usepackage{lipsum}
\usepackage{psfrag}
\usepackage{comment}
\usepackage{xcolor}
\usepackage{tikz}
\usepackage{pgfplots}
\pgfplotsset{compat=1.15}
\usepackage{mathrsfs}
\usepackage{amsmath}
\usepackage{epstopdf}
\usepackage{listings}
\usepackage{verbatim}
\usepackage{amssymb}
\usepackage{stmaryrd}
\usepackage{enumitem}
\usepackage{multirow}
\usepackage{algorithm}
\usepackage[noend]{algpseudocode}
\usepackage{cite}
\usepackage[official]{eurosym}
\usepackage{stackengine}
\usepackage{amsmath}
\usepackage{amssymb}
\usepackage{xspace}
\usepackage{stackengine}
\usepackage{verbatim}

\newtheorem{proposition}{Proposition}
\newtheorem{remark}{Remark}

\begin{document}
\IEEEoverridecommandlockouts

\title{Cramér-Rao Bounds for Integrated Sensing and Communications in  Pinching-Antenna Systems}
\author{Dimitrios Bozanis\IEEEauthorrefmark{1}, \IEEEauthorblockN{Vasilis K. Papanikolaou\IEEEauthorrefmark{2}, Sotiris A. Tegos\IEEEauthorrefmark{1}, George K. Karagiannidis\IEEEauthorrefmark{1}
 }
\IEEEauthorblockA{\IEEEauthorrefmark{1}Department of Electrical
and Computer Engineering, Aristotle University of Thessaloniki, Thessaloniki, Greece}
\IEEEauthorblockA{\IEEEauthorrefmark{2}Institute for Digital Communications (IDC), Friedrich-Alexander-University Erlangen-Nuremberg (FAU), Erlangen, Germany 
}
\IEEEauthorblockA{E-mails: dimimpoz@ece.auth.gr, vasilis.papanikolaou@fau.de, tegosoti@auth.gr, geokarag@auth.gr} 
\vspace{-0.2in}
}

\maketitle

\begin{abstract} 
Pinching-antenna systems (PASs) have recently emerged as a flexible, cost-effective route to large-scale antenna deployments envisioned for integrated sensing and communications (ISAC). This paper establishes the fundamental sensing limits of a bistatic PAS link by deriving closed-form Cramér-Rao lower bounds for the joint estimation of range and direction when a target is illuminated by pinching antennas placed along a dielectric waveguide and observed by a uniform linear array receiver. By rigorously preserving the amplitude and phase variations of each pinching antenna, as well as exploiting their non-uniform deployment, we gain valuable insights into the performance gain of PASs over conventional antenna arrays. Numerical results validate that the PAS-based ISAC can achieve centimeter-level ranging and sub-degree angular resolution with significantly fewer hardware resources than conventional uniform linear arrays. The derived bounds provide practical design guidelines for next-generation PAS-enabled ISAC systems.
\end{abstract}
\begin{IEEEkeywords}
Pinching-antenna system (PAS), Integrated Sensing and Communications (ISAC), Cramér-Rao bound, sensing accuracy.
\end{IEEEkeywords}

\section{Introduction}

Wireless communication systems have undergone significant advances in recent decades, driven by the ongoing demand for higher data rates, improved reliability, and robust security. Among the enabling technologies, multiple-input multiple-output (MIMO) systems have been instrumental in improving performance by leveraging spatial degrees of freedom (DoFs), facilitating beamforming, and significantly increasing spectral efficiency \cite{MIMO2}. Despite these benefits, conventional MIMO architectures are constrained by static antenna configurations that limit their adaptability to dynamic propagation conditions such as user mobility, environmental obstructions, and evolving network requirements. To resolve these issues, the concept of dynamic wireless channel reconfiguration has gained significant attention, enabled by emerging technologies such as reconfigurable intelligent surfaces (RISs) \cite{RIS}, movable antennas \cite{MOVE}, and fluid antennas \cite{fluid}. While these approaches offer notable advantages, they remain constrained by several limitations, including limited reconfiguration flexibility and spatial mobility, and persistent challenges in mitigating severe large-scale path loss and line-of-sight (LoS) obstructions.

To address the limitations of existing reconfigurable architectures, pinching-antenna systems (PASs) have emerged as a compelling solution for enabling highly flexible antenna configurations. First introduced by NTT DOCOMO \cite{DOCOMO}, PASs utilize low-loss dielectric waveguides populated with small dielectric elements known as pinching antennas (PAs), which can be dynamically activated and repositioned along the waveguide. This architecture unlocks significant reconfiguration capabilities, allowing on-demand adjustment of antenna positions to establish robust line-of-sight (LoS) links, mitigate large-scale path loss, and enhance spatial channel adaptability with minimal hardware complexity and cost \cite{10909665}. Recent advances in PASs have further expanded their potential by addressing a plethora of key challenges in the design, control, and optimization of wireless communication systems. In more detail, \cite{ding} demonstrated the effectiveness of PASs in strengthening LoS connectivity and adaptively overcoming severe propagation losses, while \cite{arxigos} proposed strategic PA placement algorithms that significantly improve downlink data throughput, and \cite{tyrovolas2025} derived closed-form expressions for the outage probability and average rate of PASs. In \cite{papanikolaou2025}, an artificial noise based beamforming scheme was developed to maximize the secrecy rate, thus enhancing the security of PASs, while \cite{karagkiozilikia} derived a Cramér-Rao bound (CRB) for positioning under certain noise assumptions, demonstrating improved localization accuracy.

With the advent of 6G networks, integrated sensing and communications (ISAC) has emerged as a key technology, where the network simultaneously delivers data and extracts high-resolution environmental information by transmitting the same waveform \cite{Christakis}. Leveraging the large bandwidth and extremely large-scale antenna arrays envisioned for use in 6G, ISAC promises centimeter-level ranging, sub-degree angular resolution, and gigabit-per-second data rates, all within a unified hardware platform. Such dual-function operation enables a variety of emerging use cases, including collaborative robotics, vehicular safety, and immersive extended-reality services, while reducing spectrum congestion and power consumption by eliminating the need for separate radar and communication links. The only existing work in the state-of-the-art that combines ISAC with PASs is \cite{zhang2025}, where the illumination power is maximized under a communication quality of service constraint.

In this work, motivated by the inherent DoFs of PASs, we derive the CRBs for both range and angle for a bistatic ISAC PAS, where the transmitter is a waveguide with PAs, while the receiver is a conventional uniform linear array (ULA). A special case for a far-field receiver is derived, while numerical results verify the superiority of optimized PASs in achieving significantly lower CRBs due to their reconfigurability capabilities.

\section{System Model}  

\begin{figure}
    \centering
    \includegraphics[width=1\columnwidth]{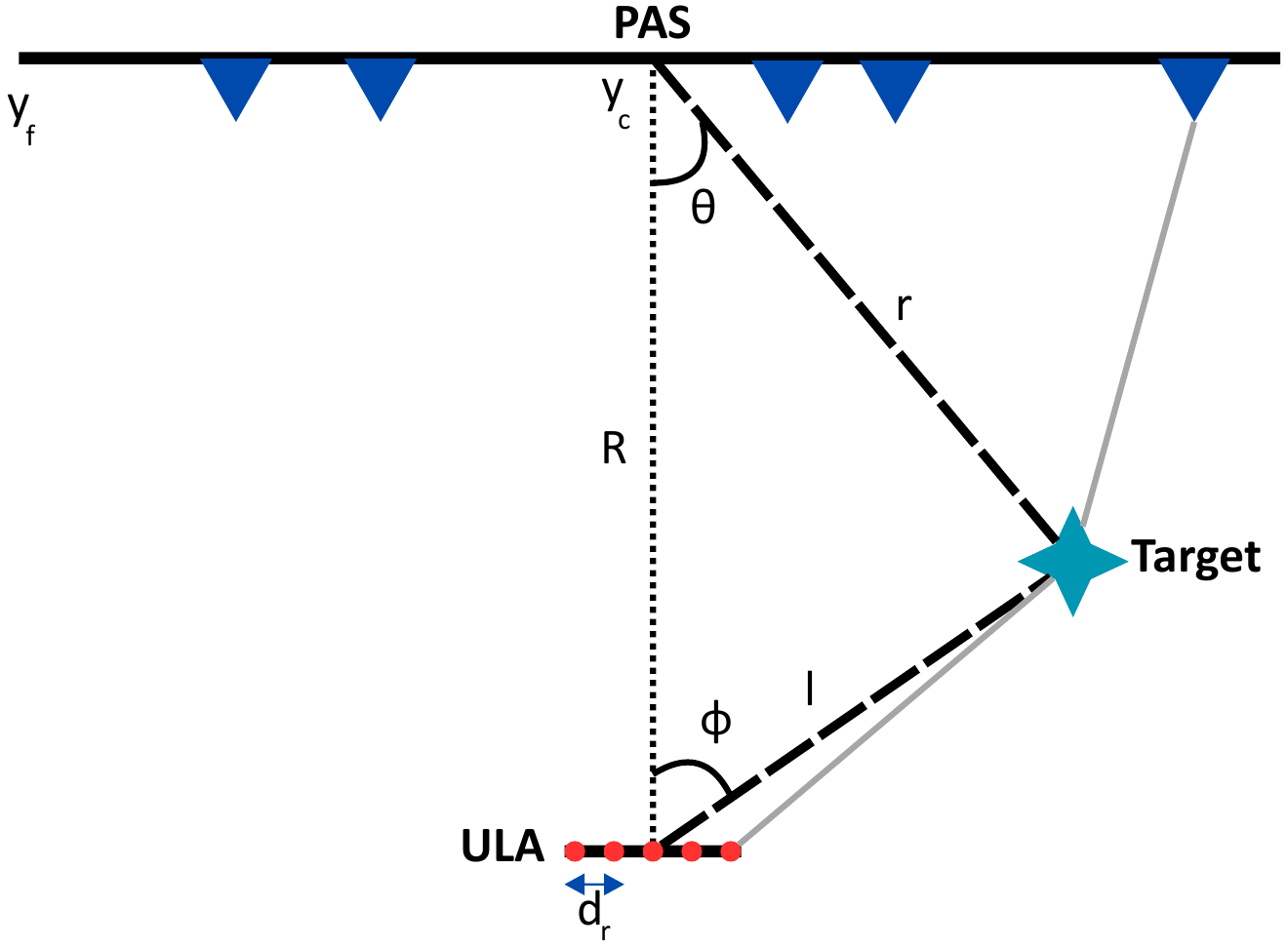}
    \caption{Illustration of a downlink PAS with multiple waveguides.}
    \label{fig:sys_model}
\end{figure}

We consider a bistatic radar-sensing configuration. The transmitter is equipped with a dielectric waveguide on which $M$ PAs are activated, and the receiver is a conventional ULA with $N$ antennas. The system model is illustrated in Fig.~\ref{fig:sys_model}. The PAs and the ULA are both deployed along the $y$-axis for $x=0$ and for $x=R$, respectively. For the set $\mathcal M\!=\{0,\pm1,\dots,\pm\frac{M-1}{2}\}$ that contains the indeces of all PAs, let $\tilde{y}_m$ denote the physical $y$-coordinates of the activated PAs. The assumption of M being an odd number is without loss of generality, since the same equations could be applied for an even number of PAs by omitting the element with index 0. Therefore, the PAs geometric center is given by
\begin{equation}
y_{\mathrm c}\triangleq\frac1M\sum_{m=-\frac{M-1}{2}}^{\frac{M-1}{2}}\tilde{y}_m,
\end{equation}
and it is chosen as the origin $(0,0)$ of the global coordinate system. Thus, the adjusted PA $y$-coordinates are given as $y_m\triangleq\tilde{y}_m-y_{\mathrm c}$, so that $\sum_{m=-\frac{M-1}{2}}^{\frac{M-1}{2}}y_m=0$. Therefore, the $m$-th PA position is given by $\mathbf w_m=[0,\;y_m]^{\mathrm T}$, where the minimum distance between two adjacent elements is $\lambda/2$, with $\lambda$ being the wavelength.

On the receiver, the center of the antenna array is placed at $[R,0]^\mathrm{T}$ and the rest of the antennas are placed with an inter-element spacing $d_R$. For the set $\mathcal N\!=\{0,\pm1,\dots,\pm\frac{N-1}{2}\}$ that contains all antennas of the ULA receiver, the $n$-th antenna position is given by $\mathbf z_n=[R,\;nd_R,]^{\mathrm T},\, n\in\mathcal N$. As in the transmitter, the assumption of an odd number of receive antennas is without loss of generality, and the analysis still holds for even values $N$.
Thus, the array apertures of the transmitter and receiver are given as
\begin{align}
 D_T&=y_{\frac{M-1}{2}}-y_{\frac{1-M}{2}},   \\
 D_R&=Nd_R,
\end{align}
respectively. The radar sensing target is characterized by range $r$ and direction $\theta$
measured from the origin, i.e., $\mathbf{q}=[r\cos\theta,\;r\sin\theta]^{\mathrm T},\,\theta\in\left(-\tfrac{\pi}{2},\tfrac{\pi}{2}\right)$.
Therefore, the distance between the target and the $m$-th PA is given by
\begin{equation}
      r_m=||\mathbf{w}_m-\mathbf{q}||=\sqrt{r^{2}-2ry_m\sin\theta+y_m^{2}},        \label{eq:r_m}
\end{equation}

In PASs, where the length of the antenna is considered to be very large, the far-field assumptions generally do not hold, so the exact distance model is needed to accurately capture the signal phase and amplitude variations over different array elements \cite{zhou2025}. Therefore, the element of the transmit array response vector depends not only on the direction $\theta$, but also on the range $r$, and can be expressed as  
\begin{equation}
      a_m(r,\theta)=
          \frac{\sqrt{\alpha_{0}}}{r_m}\;
          e^{-j \left(\frac{2\pi}{\lambda}r_m+\frac{2\pi}{\lambda_g}(y_m-y_f)\right)},               \label{eq:a_m}
\end{equation}
where $y_f$ is the $y$-coordinate of the feed point of the waveguide and $\alpha_0$ denotes the channel power gain at the reference distance of 1 m. Thus, the steering vector can be expressed as
\begin{equation}
    \mathbf a(r,\theta)=\left[a_{\frac{1-M}{2}}(r,\theta),...,a_m(r,\theta),...,a_{\frac{M-1}{2}}(r,\theta)\right],
\end{equation}
since $r_m$ is a function of $r$.

Similarly, let $l$ denote the distance between the target and the center of the receive antenna array, and $\phi$ denote the direction of the target with respect to the normal vector of the receive array. Therefore,
the element of the receive array response vector can be expressed as 
\begin{equation}
b_n(l,\phi)= \frac{\sqrt{b_{0}}}{l_n}e^{-j\frac{2\pi}{\lambda}l_n},n\in\mathcal{N},
\end{equation}
with $b_0$ denoting the channel power gain at the reference distance of 1 m and $l_n$ is given as
\begin{equation}
    l_n=||\mathbf{z}_n-\mathbf{q}||=l\sqrt{1-2n\frac{d_R}{l}\sin\phi+n^2\frac{d_R^2}{l^2}}
\end{equation}
denoting the distance between the target and the $n$-th receive antenna. However, when the distance $R$ between the transmit and receive arrays is known, the receiver-side range and angle parameters $l$ and $\phi$ can be expressed in terms of the transmitter-side parameters $r$ and $\theta$ to reduce the number of parameters to be estimated, i.e.,
\begin{equation}
    l(r,\theta)=\sqrt{R^2+r^2-2Rr\cos\theta}
\end{equation}
and
\begin{equation}
    \phi(r,\theta)=\arcsin\left(\frac{r\sin\theta}{\sqrt{R^2+r^2-2Rr\cos\theta}}\right).
\end{equation}
As a result, the distance $l_n$ and the element of the receive array response vector $b_n(l,\phi)$ can be represented in terms of $r$ and $\theta$ as
\begin{equation}\label{8}
    l_n(r,\theta) = \sqrt{R^2+r^2-2Rr\cos\theta-2nd_Rr\sin\theta+n^2d_R^2},
\end{equation}
\begin{equation}\label{9}
b_n(r,\theta)=\frac{\sqrt{b_{0}}}{l_n(r,\theta)}e^{-j\frac{2\pi}{\lambda}l_n(r,\theta).} 
\end{equation}
Therefore, the receive array response vector can be written as 
\begin{equation}
    \mathbf{b}(r,\theta)=\left[b_{-\frac{N-1}{2}}(r,\theta),...,b_n(r,\theta),...,b_{\frac{N-1}{2}}(r,\theta)\right].
\end{equation}

Note that we have expressed the receive array response vector in terms of the transmitter side angle and range parameters $(r,\theta)$ based on \eqref{8} and \eqref{9}. Thus, the received signal by the $ n$-th receive antenna due to the target reflection is then given by
\begin{equation}
  h_n(t)
  = \kappa\,b_n(r,\theta)
    \sum_{m=-\frac{M-1}{2}}^{\frac{M-1}{2}}
      a_m(r,\theta)\,x_m\!\bigl(t-\tau\bigr)
    + n_n(t),
\end{equation}
where $\kappa$ is a complex coefficient that includes the effect of the radar cross section (RCS) of the target, $\tau$ is the proportional delay of the reflected signal by the target. Here we assume that the propagation delays between different transmit and receive elements are approximately equal, which is true if $D_T + D_R \le \tfrac{c}{B}$,
where $B$ denotes the system bandwidth and $c$ is the speed of light.
Furthermore, $n_n(t)$ is the independent and identically distributed (i.i.d.) additive white Gaussian noise (AWGN) with zero mean and power spectral density $N_0$.
For ease of exposition, we consider a clutter-free model. In general, the clutter echo can be effectively mitigated by using standard clutter cancellation techniques \cite{CRB_NF}. Therefore, the vector form of the received bistatic radar signal can be written as 
\begin{equation}\label{receive}
    \mathbf{h}(t)=\kappa\mathbf{b}(r,\theta)\mathbf{a}^T(r,\theta)\mathbf{x}(t-t) + \mathbf{n}(t),
\end{equation}
where $\mathbf{x}(t)=[x_m(t)]_{m\in\mathcal{M}}$ denotes the transmitted waveform vector, $\mathbf{n}(t)\in\mathbb{C}^{Nx1}$ is the i.i.d. AWGN with zero mean and power spectral density $N_0$. 

For the sensing the transmitter forms a beam to search or track 
a target located at direction~$\theta'$ and range~$r'$. 
Therefore, the transmitted signal vector in~\eqref{receive} is given by 
\begin{equation}\label{eq:13}
\mathbf{x}(t)=
\sqrt{\frac{P}{\lVert\mathbf{a}(r,\theta)\rVert}}\,
\mathbf{a}^{*}(r',\theta')\,s(t),
\end{equation}%
where $P$ is the total transmit power, $s(t)$ is a single probing waveform satisfying
$\tfrac{1}{T_{p}}\int_{T_{p}} s(t)s^{*}(t-\alpha)\,\mathrm{d}t = R(\alpha)$, where $R(\alpha)$ is the autocorrelation function of $s(t)$.
Substituting~\eqref{eq:13} into~\eqref{receive}, the received echo can be expressed as
\begin{equation}\label{eq:14}
\mathbf{h}(t)=
\kappa\,\sqrt{\frac{P}{\lVert\mathbf{a}(r,\theta)\rVert}}\,
\mathbf{b}(r,\theta)\,\mathbf{a}^{\mathrm T}(r,\theta)
\mathbf{a}^{*}(r',\theta')\,s(t-\tau)+\mathbf{n}(t).
\end{equation}%
By applying matched filtering for $\mathbf{h}(t)$ with the transmitted waveform $s(t-a)$, and substituting $\mathbf{h}(t)$ from equation \eqref{eq:14}, we get
\begin{align}\label{eq:15}
&y(\alpha,r',\theta') =
\frac{1}{\sqrt{T_{p}}}\int_{T_{p}} \mathbf{h}(t)\,s^{*}(t-\alpha)\,\mathrm{d}t
\\
&=\kappa\,\sqrt{\frac{T_{p}P}{\lVert\mathbf{a}(r,\theta)\rVert}}\,
\mathbf{b}(r,\theta)\,\mathbf{a}^{\mathrm T}(r,\theta)
\mathbf{a}^{*}(r',\theta')\,R(\alpha-\tau)+\mathbf{\tilde{n}}, \nonumber
\end{align}%
where $\tilde{\mathbf n}\triangleq\frac{1}{\sqrt{T_p}}\displaystyle\int_{T_p} n(t-\alpha)\,s^{*}(t)\,dt$ is the resulting noise vector with zero mean and variance $N_0$. When the search parameters match the true target parameters,
$\alpha=\tau$, $r'=r$, and $\theta'=\theta$, and by setting $S(r,\theta) \triangleq \sum_{m} a_{m}(r,\theta)$, the observation simplifies to
\begin{equation}\label{eq:16}
\mathbf{y}= \kappa\sqrt{T_{p} P}
S(r,\theta)\mathbf{b}(r,\theta)+\tilde{\mathbf n}.
\end{equation}%
Having developed the bistatic PAS-based ISAC system model and distilled the matched-filter output into the compact form of \eqref{eq:16}, we now leverage this formulation in Section \ref{sec:CRB} to derive closed-form CRBs that characterize the fundamental limits of joint range and angle estimation. 
\section{Cramér–Rao Bound for PASs}\label{sec:CRB}
In this section, we establish the fundamental estimation limits of the proposed PA radar by deriving closed-form CRBs for the target range~\(r\) and direction~\(\theta\).
The CRB represents the minimum achievable mean-square error of any unbiased estimator and thus serves as a key sensing performance benchmark \cite{Christakis}.
Starting with the matched-filter observation in~\eqref{eq:16}, we construct the Fisher information matrix, isolate the sub-matrix associated with \((r,\theta)\), and obtain explicit formulas for the CRBs for range and angle estimation \(\mathrm{CRB}_{r}\) and \(\mathrm{CRB}_{\theta}\). These expressions demonstrate the influence of PA placement, receive array aperture, and carrier wavelength on the sensing accuracy.
For ease of notation, we start by setting 
\begin{align}
   \rho &\triangleq \sqrt{T_p {P}} \kappa, \\
   \mathbf{g}(r, \theta) &\triangleq S(r, \theta) \, \mathbf{b}(r, \theta), \label{eq:g}
\end{align}
and thus \eqref{eq:16} can be written as 
\begin{equation}\label{eq:17}
\mathbf{y}= \rho\,\mathbf{g}+\tilde{\mathbf n}.
\end{equation}%

\begin{proposition}\label{l1}
The CRBs for range and angle estimation are given respectively in closed form as 
\begin{align} \label{eq:crbtheta}
\mathrm{CRB}_\theta &= \left[\mathbf{Q}^{-1}\right]_{11} \frac{\sigma^2}{2L} 
= \frac{\sigma^2}{2L |\rho|^2}   \frac{s}{is - k^2}, \\\label{eq:crbr}
\mathrm{CRB}_r &= \left[\mathbf{Q}^{-1}\right]_{22} \frac{\sigma^2}{2L} 
= \frac{\sigma^2}{2L |\rho|^2}   \frac{i}{is - k^2},
\end{align}
where $L\triangleq  BT_p$ denotes the time-bandwidth in a CPI and $\sigma^2=N_0B$ and
\begin{equation}
\begin{split}
i &= \|\mathbf{g}_\theta\|^2 - \frac{|\mathbf{g}_\theta^H \mathbf{g}|^2}{\|\mathbf{g}\|^2}, \\
s &= \|\mathbf{g}_r\|^2 - \frac{|\mathbf{g}_r^H \mathbf{g}|^2}{\|\mathbf{g}\|^2},\label{CRB} \\
k &= \Re\left\{ \mathbf{g}_\theta^H \mathbf{g}_r \right\}
- \frac{\Re\left\{ \mathbf{g}_\theta^H \mathbf{g} \right\} \Re\left\{ \mathbf{g}_r^H \mathbf{g} \right\}}{\|\mathbf{g}\|^2},
\end{split}
\end{equation}
with $\mathbf{g}_\theta = \frac{\partial \mathbf{g}}{\partial \theta}$ and $
\mathbf{g}_r = \frac{\partial \mathbf{g}}{\partial r}$.
\end{proposition}
\begin{IEEEproof}
The proof is presented in Appendix A.
\end{IEEEproof}
With the general CRB formulas established, we now analyze how the receiver geometry influences these bounds.

\begin{proposition}\label{l2}
For a plane wave receiver, the CRB expressions for both range and angle estimation degenerate to
\begin{align}
    \mathrm{CRB}_{r}\rightarrow+\infty , \\\mathrm{CRB}_{\theta}\rightarrow+\infty.
\end{align}
\end{proposition}
\begin{IEEEproof}
If $l\gg 1.2D_R$, the receiver can be treated as a plane wave receiver \cite{CRB_NF}, thus the amplitude variations in the receiver can be neglected since $1/l_n\approx1/l$ and the phase term can also be approximated as $l_n\approx l$. Thus, $\mathbf g(r,\theta)$ can be expressed as
\begin{equation}
   \mathbf g(r,\theta)=S(r,\theta)\,
                      e^{-j\frac{2\pi}{\lambda}l(r,\theta)}
                      \mathbf{1}_{N},
\end{equation}
where $\mathbf{1}_{N}$ is the all-ones vector of size $N$. Since $\mathbf 1_{N}$ is constant, $\mathbf g_r$ and $\mathbf g_\theta$ are simplified to
\begin{align}
   \mathbf g_r &= S_r\mathbf b
               -j\frac{2\pi}{\lambda}l_r\,S\mathbf b\\
   \mathbf g_\theta &= S_\theta\mathbf b
               -j\frac{2\pi}{\lambda}l_\theta\,S\mathbf b.
               \label{eq:gt_colinear}
\end{align}
Thus, we observe that $\mathbf g(r,\theta)$, $\mathbf g_r$, $\mathbf g_\theta$ are colinear, since the coefficients of $\mathbf b$ are scalars. Therefore, by setting $\hat{\mathbf{b}}=\mathbf{b}/\sqrt{N}$ and 
\begin{equation}
\begin{aligned}
        \beta        &= S\sqrt N,\\
        \beta_r      &= (S_r-j\frac{2\pi}{\lambda}l_r\,S)\sqrt N,\\
        \beta_\theta &= (S_\theta-j\frac{2\pi}{\lambda}l_\theta\,S)\sqrt N,
\end{aligned}
\end{equation}
the vector norms can be rewritten as
\begin{equation}
\begin{aligned}
   \|\mathbf g\|^{2}         &= |\beta|^{2},\\
   \|\mathbf g_r\|^{2}       &= |\beta_r|^{2},\\
   \|\mathbf g_\theta\|^{2}  &= |\beta_\theta|^{2}.
\end{aligned}
\end{equation}
Finally, $s$, $i$ and $k$ can be calculated as 
\begin{align}
   s &= \|\mathbf g_r\|^{2}
        -\frac{|\mathbf g_r^{H}\mathbf g|^{2}}{\|\mathbf g\|^{2}}
      = |\beta_r|^{2}-|\beta_r|^{2}
      = 0, \\[4pt]
   i &= \|\mathbf g_\theta\|^{2}
        -\frac{|\mathbf g_\theta^{H}\mathbf g|^{2}}{\|\mathbf g\|^{2}}
      = |\beta_\theta|^{2}-|\beta_\theta|^{2}
      = 0, \\[4pt]
   k &= \Re\{\mathbf g_\theta^{H}\mathbf g_r\}
        -\frac{\Re\{\mathbf g_\theta^{H}\mathbf g\}
                \Re\{\mathbf g_r^{H}\mathbf g\}}
               {\|\mathbf g\|^{2}}
      = 0,
\end{align}
which results in both CRB expressions to be infinite, i.e., $\mathrm{CRB}_{r}\rightarrow+\infty$ and $\mathrm{CRB}_{\theta}\rightarrow+\infty$, which concludes the proof.
\end{IEEEproof}

\begin{remark}
The joint divergence of $\mathrm{CRB}_{r}$ and $\mathrm{CRB}_{\theta}$ for a plane wave receiver arises from the fact that $\mathbf b(r,\theta)$ collapses to a rank-one vector $e^{-j 2\pi l/\lambda}\mathbf 1_N$ once amplitude variations and spherical wave phases are neglected.
However, in the case of a PAS receiver, this degeneracy cannot occur, even if the distance between the target and the receiver is considerable. Since the waveguide in which the PAs are deployed is much larger than a conventional ULA, the condition $l\gg 1.2D_R$ can still be unsatisfied even at ranges of hundreds of meters. Thus, the large geometric differences $l_n-l$ increase the FIM due to significant phase and amplitude variations between the PAs. Furthermore, each PA contributes an additional, range-independent, in-waveguide phase term ${e^{-j2\pi(y_n-y_f)/\lambda_g}}$, which guarantees that the vectors $\mathbf g,\mathbf g_r,\mathbf g_\theta$ are no longer colinear. Thus, the Schur complement determinant $is-k^2$ is strictly positive, and both CRBs remain finite even for large values of $l$.
Consequently, replacing the conventional ULA with a PAS receiver preserves the amplitude and phase diversity required for well-conditioned estimation, so that the theoretical bounds remain finite for any practical target distance.
\end{remark}

\section{Numerical Results}
In this section, numerical results are provided to validate our derived CRB expressions. Monte Carlo simulations are performed where the target angle and range follow a uniform distribution with $\theta \in[-\pi/6,\pi/6]$ and $r\in[5,25]$. The transmit and receive array distance is set to $R=30$ m and the transmit power is set to $P=0$ dBm, while the carrier frequency is $f=27$ GHz. The length of the waveguide is set to $D_T=10$ m, the noise power is set to $\sigma^2=-90$ dBm, while the effective refractive index of each dielectric waveguide is set to $n_\mathrm{eff}=1.4$. The positions of the PAs are optimally chosen by a global search algorithm to minimize the CRBs, further reducing the achievable bounds.

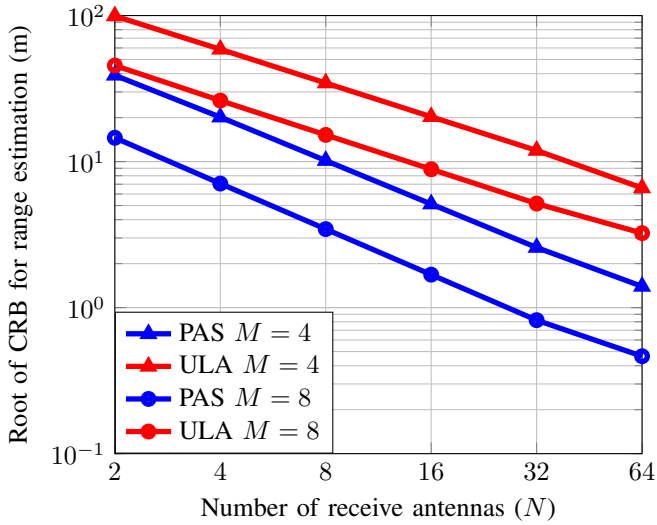
\begin{figure}
\centering
\begin{tikzpicture}
\begin{loglogaxis}[
    width=0.97\linewidth,
    xlabel={Number of receive antennas ($N$)},
    ylabel={Root of CRB for range estimation (m) },
    legend columns=1,
    legend cell align = {left},
    legend style={at={(0,0)}, anchor=south west},
    grid=both,
     xmin=2, xmax=64,
    ymin=1e-1, ymax=1e2,
    xtick={2,4,8,16,32,64},
    xticklabels={$2$,$4$,$8$,$16$,$32$,$64$}
]

\addplot+[
            blue,
            mark=triangle,
            mark repeat = 1,
            mark size = 2,
            line width = 2pt,
            style = solid,] table {PAS_4.dat};
\addlegendentry{PAS $M=4$}

\addplot+[
            red,
            mark=triangle,
            mark repeat = 1,
            mark size = 2,
            line width = 2pt,
            style = solid,] table {ULA_4.dat};
\addlegendentry{ULA $M=4$}

\addplot+[
            blue,
            mark=o,
            mark repeat = 1,
            mark size = 2,
            line width = 2pt,
            style = solid,] table {PAS_8.dat};
\addlegendentry{PAS $M=8$}

\addplot+[
            red,
            mark=o,
            mark repeat = 1,
            mark size = 2,
            line width = 2pt,
            style = solid,] table {ULA_8.dat};
\addlegendentry{ULA $M=8$}

\end{loglogaxis}
\end{tikzpicture}
\caption{\centering CRB for range estimation vs the number of receive antennas ($N$).}
\label{fig:range}
\end{figure}

In Fig. \ref{fig:range}, the square root of the CRB for range estimation is plotted against the number of receive antennas~$N$ for different PAS and ULA transmitters with $M=4$ and $M=8$ antennas. The CRB decreases almost exactly with slope~$-1$, in agreement with the analytical result
$\sqrt{\mathrm{CRB}_{r}}\propto N^{-1}$ derived from the FIM. Thus, doubling $N$ halves the minimum achievable standard deviation of any unbiased range estimator. Doubling the number of transmit antennas also significantly improves performance due to the expected beamforming gain.
However, the most valuable insight of the figure is the systematic gap between the optimized PAS and the conventional ULA, since the CRB experiences an immense reduction of the bound over the entire $N$ range, where even the PAS with $M=4$ outperforms an ULA with $M=8$.
This gain is produced by two diversity mechanisms that are absent in a conventional ULA, the first being that the non-uniform antenna positions of the PAs increase the effective aperture and steepen the phase gradients $\partial a_{m}/\partial(r,\theta)$, and the second being that the phase term $2\pi (y_{m}-y_{f})/\lambda_{g}$ in the waveguide provides an additional DoF to optimize accordingly.
Both effects increase the determinant $is-k^{2}$ of the information and thus decrease the CRB.

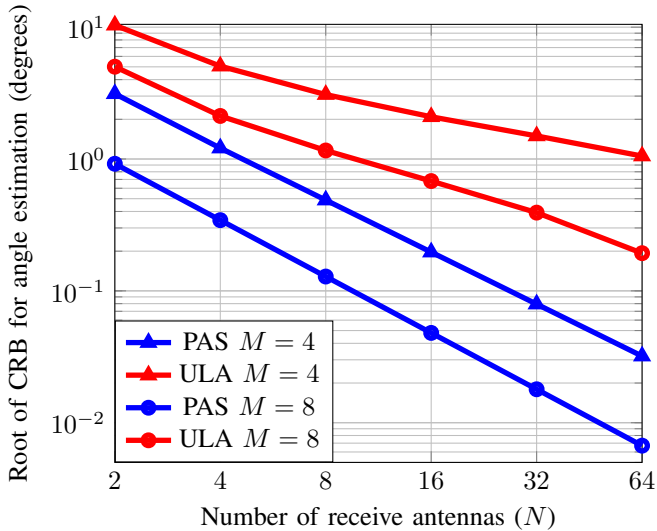
\begin{figure}
\centering
\begin{tikzpicture}
\begin{loglogaxis}[
    width=0.97\linewidth,
    xlabel={Number of receive antennas ($N$)},
    ylabel={{Root of CRB for angle estimation (degrees) }},
    xmin=2, xmax=64,
    ymin=5e-3, ymax=10.5,
    xtick={2,4,8,16,32,64},
    xticklabels={$2$,$4$,$8$,$16$,$32$,$64$},
    grid=both,
    legend style={at={(0,0)}, anchor=south west},
]
\addplot+[
            blue,
            mark=triangle,
            mark repeat = 1,
            mark size = 2,
            line width = 2pt,
            style = solid,] table {Pas_4.dat};
\addlegendentry{PAS $M=4$}

\addplot+[
            red,
            mark=triangle,
            mark repeat = 1,
            mark size = 2,
            line width = 2pt,
            style = solid,] table {Ula_4.dat};
\addlegendentry{ULA $M=4$}

\addplot+[
            blue,
            mark=o,
            mark repeat = 1,
            mark size = 2,
            line width = 2pt,
            style = solid,] table {Pas_8.dat};
\addlegendentry{PAS $M=8$}

\addplot+[
            red,
            mark=o,
            mark repeat = 1,
            mark size = 2,
            line width = 2pt,
            style = solid,] table {Ula_8.dat};
\addlegendentry{ULA $M=8$}

\end{loglogaxis}
\end{tikzpicture}
\caption{\centering CRB for angle estimation vs the number of receive antennas ($N$).}
\label{fig:angle}
\end{figure}

Similarly, in Fig. \ref{fig:angle}, the square root of the CRB for angle estimation under identical conditions is plotted for different numbers of receive antennas for the same transmitter configurations as in Fig. \ref{fig:range}. The same $N^{-1}$ slope is observed, but the absolute differences between the different architectures are more pronounced, as the accuracy of angle estimation is significantly improved by spatial aperture and phase diversity.
With $M=8$ and $N=32$, the PAS achieves a standard deviation below $0.01^\circ$, about an order of magnitude better than the ULA baseline.
Moreover, the doubling of $M$ is significantly more valuable for angle estimation than for range, as the coherent beamforming gain is amplified by the additional leverage that widely spaced PAs exert on the angular phase gradient, resulting in an almost $10$-fold reduction in the square root of the CRB compared to the $M=4$ ULA with $N\ge 16$. Thus, both figures demonstrate that PAS can not only improve the received SNR, but also provide the required diversity at the transmitter to reduce both the range and angle CRBs, providing ultra-precise range and angle estimations and a remarkable gain over conventional MIMO systems.

\section{Conclusions}
In this work, we derived the fundamental sensing limits of a bistatic ISAC system with a PAS as a transmitter and a ULA receiver, by providing closed-form CRBs for joint range and direction estimation. The presented analysis revealed three key findings. First, the non-uniform placement of the PAs, as well as the additional phase term due to in-waveguide propagation, provides valuable spatial-phase diversity that sharply tightens the bounds relative to a uniform linear array. In addition, the resulting centimeter-level ranging and sub-degree angular resolution offered by the proposed system, with far fewer front-end components than conventional architectures, provides a cost-effective alternative to the extremely large arrays envisioned for 6G. Finally, the impact of amplitude and phase diversity between the antenna elements was verified in the proposed system as the bounds alternatively diverge. Overall, the current framework sets a precedent for the capabilities offered by PAS-enabled ISAC and invites future work with more complex propagation, multi-target schemes, and hardware impairments.
\appendices

\section*{Appendix A - Proof of Proposition \ref{l1}}\label{ap:ferrari}
Let $\mathbf{w}=\rho\mathbf{g}$ and define the real parameter vector
$\boldsymbol{z}=[\,\theta,\; r,\; \kappa_{\mathrm r},\; \kappa_{\mathrm
i}\,]^{\mathsf T}$, where $\kappa_{\mathrm r}$ and
$\kappa_{\mathrm i}$ are the real and imaginary parts of~$\kappa$, respectively.  
The Fisher information matrix (FIM) for~$\boldsymbol{z}$ is given by
\begin{equation*}\label{eq:FIM}
\mathbf{F} = \frac{2}{N_0} \, \Re \left\{ 
\left( \frac{\partial \mathbf{w}}{\partial \mathbf{z}} \right)
\left( \frac{\partial \mathbf{w}}{\partial \mathbf{z}} \right)^H 
\right\}
\end{equation*}
\begin{align}
= \frac{2}{N_0}
\left[
\begin{array}{cc|cc}
\upsilon_{\theta\theta} & \upsilon_{\theta r} & \upsilon_{\theta \kappa_r} & \upsilon_{\theta \kappa_i} \\
\upsilon_{\theta r} & \upsilon_{rr} & \upsilon_{r \kappa_r} & \upsilon_{r \kappa_i} \\ \hline
\upsilon_{\theta \kappa_r} & \upsilon_{r \kappa_r} & \upsilon_{\kappa_r \kappa_r} & 0 \\
\upsilon_{\theta \kappa_i} & \upsilon_{r \kappa_i} & 0 & \upsilon_{\kappa_i \kappa_i}
\end{array}
\right] =
\left[
\begin{array}{c|c}
\bm{\Pi}_{11} & \bm{\Pi}_{12} \\ \hline
\bm{\Pi}_{12}^{\mathrm{T}} & \bm{\Pi}_{22}
\end{array}
\right],
\end{align}
where $\upsilon_{z_1 z_2} \triangleq \Re \left\{ \left( \frac{\partial \mathbf{w}}{\partial z_1} \right) \left( \frac{\partial \mathbf{w}}{\partial z_2} \right)^H \right\}$. Note that $\kappa_i$ and $\kappa_r$ are unknown auxiliary parameters that are independent of angle and range and include the effect of RCS of the target. The CRB of the parameters of interest $(r,\theta)$ is related to the inverse of FIM as \cite{CRB_NF}
\begin{align}
\mathbf{F}^{-1} = \frac{N_0}{2}
\left[
\begin{array}{c|c}
\mathbf{Q}^{-1} & \times \\
\hline
\times & \times
\end{array}
\right],
\end{align}
where
$\mathbf{Q} = \bm{\Pi}_{11} - \bm{\Pi}_{12} \bm{\Pi}_{22}^{-1} \bm{\Pi}_{12}^{T}$ is the Schur complement of $\bm{\Pi}_{22}$ corresponding to $\mathbf{F}$. Thus, after some mathematical manipulations, we derive that 
\begin{align}
\mathbf{Q} &= |\rho|^2 \left[
\left(
\begin{array}{cc}
\|\mathbf{g}_\theta\|^2 & \Re\left\{ \mathbf{g}_\theta^H \mathbf{g}_r \right\} \\
\Re\left\{ \mathbf{g}_\theta^H \mathbf{g}_r \right\} & \|\mathbf{g}_r\|^2
\end{array}
\right) \right. \nonumber \\
&\left.
- \frac{1}{\|\mathbf{g}\|^2}
\left(
\begin{array}{cc}
|\mathbf{g}_\theta^H \mathbf{g}|^2 & \Re\left\{ \mathbf{g}_\theta^H \mathbf{g} \right\} \Re\left\{ \mathbf{g}_r^H \mathbf{g} \right\} \\
\Re\left\{ \mathbf{g}_\theta^H \mathbf{g} \right\} \Re\left\{ \mathbf{g}_r^H \mathbf{g} \right\} & |\mathbf{g}_r^H \mathbf{g}|^2
\end{array}
\right)
\right],
\end{align}
where $\mathbf{g}_\theta = \frac{\partial \mathbf{g}}{\partial \theta}$ and $
\mathbf{g}_r = \frac{\partial \mathbf{g}}{\partial r}$. By defining $i$,$s$,$k$ as in equation \eqref{CRB}, the matrix $\mathbf{Q}$ could be rewritten as 
\begin{equation}
\mathbf{Q} = |\rho|^2 
\begin{pmatrix}
i & k \\
k & s
\end{pmatrix}
\end{equation}
and thus, the arising CRBs for both angle and range estimation are presented in equations \eqref{eq:crbtheta} and \eqref{eq:crbr}, respectively. To derive the closed-form expressions of both CRBs, the terms of $i$, $s$, $k$ need to be also calculated. Therefore, in the following, we analytically calculate each term found in \eqref{CRB}. By applying the chain derivation rule to $\mathbf{g}$, as it was defined for \eqref{eq:g}, with regards to $r$ and $\theta$, we get
\begin{equation}\label{eq:gtheta}
    \mathbf{g}_\theta = S_\theta\mathbf{b}+S\mathbf{b}_\theta
\end{equation}
and
\begin{equation}\label{eq:gr}
    \mathbf{g}_r = S_r\mathbf{b}+S\mathbf{b}_r,
\end{equation}
respectively. Similarly, for the transmitter side, we can express $S_\theta$ and $S_r$ as
\begin{align}
S_\theta &= \sum_m \frac{\partial a_m}{\partial \theta}, \\
S_r &= \sum_m \frac{\partial a_m}{\partial r},
\end{align}
respectively, with
\begin{align}
\frac{\partial a_m}{\partial \theta} &= a_m \left[ -\frac{1}{r_m} \frac{\partial r_m}{\partial \theta} 
- j \frac{2\pi}{\lambda} \frac{\partial r_m}{\partial \theta} \right], \\
\frac{\partial a_m}{\partial r} &= a_m \left[ -\frac{1}{r_m} \frac{\partial r_m}{\partial r} 
- j \frac{2\pi}{\lambda} \frac{\partial r_m}{\partial r} \right],
\end{align}
and 
\begin{align}
\frac{\partial r_m}{\partial \theta} &= -\frac{r\, y_m \cos \theta}{r_m}, \\
\frac{\partial r_m}{\partial r} &= \frac{r - y_m \sin \theta}{r_m}.
\end{align}
From the receiver steering vectors for the $n$-th receive antenna, we have
\begin{align}
({b}_\theta)_n &= \frac{\partial b_n}{\partial \theta}, \\
({b}_r)_n &= \frac{\partial b_n}{\partial r},
\end{align}
where
\begin{align}
\frac{\partial b_n}{\partial \theta} &= b_n \left[ -\frac{1}{l_n} \frac{\partial l_n}{\partial \theta} 
- j \frac{2\pi}{\lambda} \frac{\partial l_n}{\partial \theta} \right], \\
\frac{\partial b_n}{\partial r} &= b_n \left[ -\frac{1}{l_n} \frac{\partial l_n}{\partial r} 
- j \frac{2\pi}{\lambda} \frac{\partial l_n}{\partial r} \right].
\end{align}
and
\begin{align}
\frac{\partial l_n}{\partial \theta} &= \frac{R r \sin \theta - n d_R r \cos \theta}{l_n}, \\
\frac{\partial l_n}{\partial r} &= \frac{r - R \cos \theta - n d_R \sin \theta}{l_n}.
\end{align}
Therefore, by taking into account the definition of $\mathbf{g}$, \eqref{eq:gtheta}, and \eqref{eq:gr}, we get
\begin{align} \label{eq:iskg}
\mathbf{g}_\theta^H \mathbf{g} &= S_\theta^* S   \|\mathbf{b}\|^2 + S^* S_\theta   \mathbf{b}_\theta^H \mathbf{b}, \\ \label{eq:iskgg}
\mathbf{g}_r^H \mathbf{g} &= S_r^* S   \|\mathbf{b}\|^2 + S^* S_r   \mathbf{b}_r^H \mathbf{b}, \\
\mathbf{g}_\theta^H \mathbf{g}_r &= S_\theta^* S_r   \|\mathbf{b}\|^2 + S^* S_r   \mathbf{b}_\theta^H \mathbf{b} \nonumber \\ \label{eq:iskggg}
&+ S_\theta^* S   \mathbf{b}^H \mathbf{b}_r + |S|^2   \mathbf{b}_\theta^H \mathbf{b}_r.
\end{align}
Thus, we can plug \eqref{eq:iskg}, \eqref{eq:iskgg}, and \eqref{eq:iskggg} into \eqref{CRB}, to calculate in closed form $i$, $s$, $k$, and finally $\mathrm{CRB}_\theta$ and $\mathrm{CRB}_r$, as in \eqref{eq:crbtheta} and \eqref{eq:crbr}, which concludes the proof.

\bibliographystyle{IEEEtran}
\bibliography{IEEEabrv,references}
\end{document}